\documentclass[10pt,letterpaper]{article}
\usepackage{opex3}

\begin{document}

\title{Fabrication of mesoscale polymeric templates for three-dimensional disordered photonic materials}

\author{Jakub Haberko$^{1,2}$ and Frank Scheffold$^{1}$ }

\address{$^1$Physics Department and Fribourg Center for Nanomaterials, University of Fribourg, Chemin de Mus\'{e}e 3, 1700 Fribourg, Switzerland}
\address{$^2$Currently with the Faculty of Physics and Applied Computer Science, AGH University of Science and Technology, al. Mickiewicza 30, 30-059 Krakow, Poland }

\email{Frank.Scheffold@unifr.ch} 
\homepage{http://physics.unifr.ch/en/page/54/} 


\begin{abstract} We report on the mesoscale fabrication and characterization of polymeric templates for isotropic photonic materials derived from hyperuniform point patterns using direct laser writing in a polymer photoresist. We study experimentally the microscopic structure by electron microscopy and small angle light scattering.   Reducing the refractive index mismatch by liquid infiltration we find good agreement  between the scattering data and numerical calculations based on a discrete dipole approximation.  Our work demonstrates the feasibility of fabricating such random designer materials on technologically relevant length scales. \end{abstract}




\section{Introduction}

Coherent scattering can strongly influence the macroscopic transport of electrons or photons. In the limit of weak scattering from periodic lattices the classical concept of Bloch allows to distinguish propagating and non-propagating wave vectors $\bf{k}$. For a periodic one-dimensional structure with a lattice constant $d$ forbidden wave vectors appear around the edge of the Brillouin zone $k =\pi/a$, the latter being a signature of back-reflection from Bragg planes with scattering vectors $q=2k=2\pi/a$ \cite{ref1}. Despite the beauty of BlochÕs theory it is well known that the model covers only a small part of actual electron transport phenomena in metals or semi-conductors. A prime example is the fact that many alloys and even liquids display metallic or semiconducting properties \cite{Wa04}. \newline The broader concept of electronic scattering from ordered or disordered structures can be readily transferred to the case of electromagnetic waves as shown in the pioneering work of Yablonovitch and John \cite{ref9,ref10}. Following their discovery there have been enormous research efforts aimed at manufacturing photonic semi-conducting materials. Again these efforts have largely concentrated on crystalline structures, owing to the simple deterministic design rules and the fact that band-structure calculations are well established for crystals \cite{ref1,ref11,ref12}. The structural anisotropy of periodic structures, both in real space and reciprocal-space, however, imposes physical limitations for three-dimensional photonic materials. A direct consequence of anisotropy is the appearance of stop bands with varying strengths depending on the direction of wave propagation. In order to achieve complete band-gaps in three dimensions the gaps in all directions have to overlap, a condition that is difficult to achieve or that requires a very large dielectric contrast. As emphasized previously \cite{ref7,ref8}, the availability of isotropic photonic materials would thus be highly desirable. A step in this direction has been made with the discovery that quasi-crystalline structures equally possess photonic properties while having a much higher degree of rotational symmetry \cite{Flor09QC,ref13}. Quasicrystalline structures are aperiodic but nevertheless exhibit sharp Bragg peaks \cite{ref14}. Since they are ordered and can be projected from higher-dimensional crystalline structures they can be considered a case in between crystalline and amorphous.
Disordered dielectric heterostructures have also been considered candidates for photonic band gap materials \cite{ref2, Rec11}. However, until recently, the design rules needed to derive such structures have remained obscure. A number of studies have looked into materials with short-range order such as photonic liquids and glasses derived from self-assembled colloidal \cite{ref3,ref4,ref5} or biomimetic materials \cite{ref6}. Although Bragg-like scattering and pseudo gaps have been observed no conceptual proof for the existence of a complete band gap could be derived even at arbitrarily high dielectric contrast. Moreover, the interplay between short range-order and Anderson localization of light in three dimension, highlighted in the early work of Sajeev John \cite{ref10}, remains an unsolved question.

Recently Florescu, Torquato and Steinhardt have taken a different route to design amorphous structures with a full photonic bandgap (PBG)\cite{ref7}. The authors make the general claim that three structural conditions have to be satisfied to obtain a PBG: hyperuniformity, uniformlocal topology, and short-range geometric order. Structures that fulfill these conditions display a vanishing structure factor $S(q)$ at small but finite wavenumbers $\left| \bf{q} \right| \le \left| {\bf{q_c}} \right|$, a structural property coined ÔstealthyÕ hyperuniformity. Above a threshold dielectric contrast such structures are predicted to display a full PBG.  A trivial finding is that all crystalline and quasicrystalline materials fall into this class. Their main point however is that they could identify peculiar structures that are disordered and thus completely isotropic but still fulfill the above mentioned conditions \cite{ref7}. It has been shown that for exampled randomly jammed packings of spheres posses such hyperuniform long-range correlations \cite{Zac11}. In addition to the underlying geometric order, the local topology plays an important role \cite{ref8}. Mapping hyperuniform point patterns with short-range geometric order into tessellations allows the design of interconnected networks that gives rise to enhanced photonic properties. Numerical calculations for such two- and three-dimensional disordered designer materials \cite{ref7,ref8} indicate the presence of a robust full band-gap in the limit of sufficiently high dielectric contrast, typically $n>3$ in air. Preliminary experimental data obtained from 2D hyperuniform structures in the microwave regime seem to confirm these predictions \cite{Man10}. 

Here we report on the first experimental realization of micron scale three-dimensional polymeric templates, n $\sim 1.52$, based on the design rules derived by Florescu and coworkers \cite{ref7}.   Our present aim is to demonstrate the feasibility of fabricating such random designer materials on technologically relevant length scales, a first major step towards the fabrication of three dimensional amorphous full PBG materials.

\begin{figure}[htbp]
\centering\includegraphics[width=11cm]{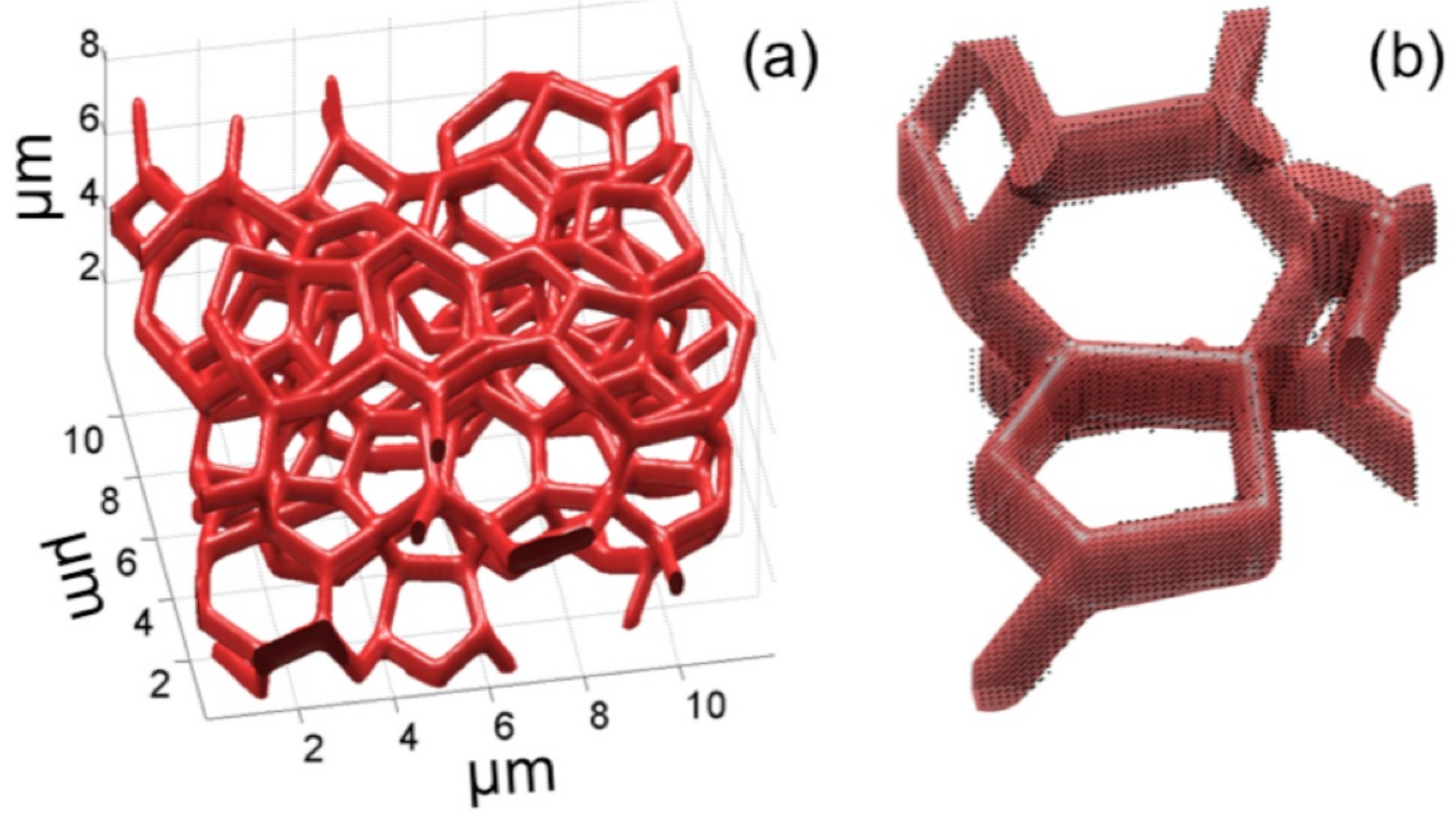}
\caption{Design of amorphous photonic structures based upon a tetrahedral network of elliptical rods. (a) Enlarged view of part of the network obtained by tessellation of a hyperuniform seed pattern, rod cross section $840 \times280$nm$^2$. (b) close-up showing also the grid points lying inside or at the surface of the rods used for the numerical analysis of the scattering pattern.}
\label{fig1}
\end{figure}

\section{Results}
\subsection{Nanofabrication of designer disordered materials in a polymer photoresist.}  
\begin{figure}[htbp]
\centering\includegraphics[width=10cm]{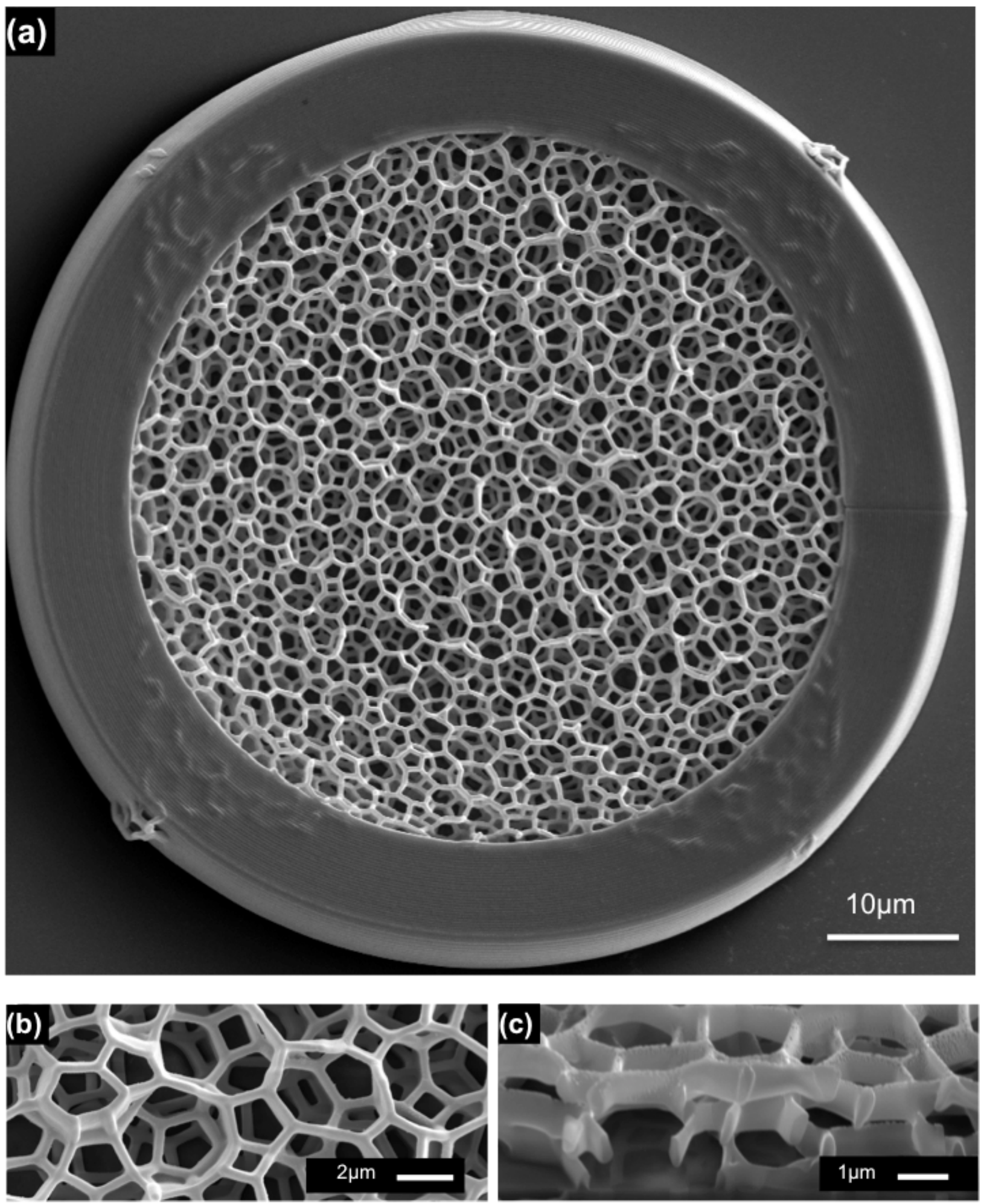}
\caption{Electron micrographs of fabricated hyperuniform three-dimensional disordered structures. (a) Normal view of a structure with height $h = 8\mu$m and inner diameter $d =65 \mu m$  (b) close-up view (c) focused ion beam cut of the same structure.}
\label{fig2}
\end{figure}

We fabricate mesoscale polymer structures by direct laser writing (DLW) (Photonic Professional, Nanoscribe, Germany).  The method allows to replicate a 3D structure voxel by voxel into a polymer photoresist with submicron resolution using a two-photon polymerization process \cite{ref12}. We first implement the protocol suggested by Florescu and coworkers to map a hyperuniform point pattern into tessellations for photonic materials design \cite{ref7}. As a seed structure we use the centroid positions from a maximally randomly jammed assembly of spheres of diameter $a = 3.31\mu$m  with a volume filling fraction of $\phi \simeq 0.64$ \cite{ref15}.  As shown by Torquato and coworkers \cite{Zac11} such structures indeed possess the required hyperuniform long-range correlations. Locally the seed pattern displays pronounced short-range correlations due to close packing of the original assembly of spheres. Therefore the real space pair correlation function $g(r)$ is sharply peaked at contact $\left| \bf{r} \right|= d$ and thus the structure factor $S(q)$ is peaked at $\left| \bf{q} \right|_{max}=2\pi/d$ \cite{Don05}.  The latter is closely related to the well-known diffraction ring observed in experiments on dense molecular and complex liquids or glasses \cite{Bar03}. Next we perform a 3D Delaunay tessellation of the spheres center positions. In this scheme tetrahedrons are formed in such a way that no sphere center is contained in the circumsphere of any tetrahedron in the tessellation. The center-of-mass of neighboring tetrahedrons are then connected resulting in a 3D random tetrahedral network with the desired hyperuniform properties. The only parameters not yet fixed are related to the shape and cross section of the dielectric rods replacing the fictitious connection lines (Fig. \ref{fig1} (a)). 
\newline We find optimal conditions in the writing process when the rods are written into IPG-780 negative tone photoresist (Nanoscribe, Germany). The laser writing pen has an elliptical cross section of about $840 \times 280$ nm$^2$. When writing thinner rods the structure becomes mechanically unstable while thicker rods lead to overfilling of the structure. The aspect ratio of ca. $3.0$ of the rods is dictated by the point spread function of the illuminating microscope objective during the writing process which is set by the refractive index of about $1.52$ of the material and the numerical aperture of $1.4$. For the parameters chosen we estimate the volume-filling fraction of the rods to be roughly 8 $\%$. \newline In comparison to the well-established case of periodic rod assemblies \cite{ref12} the fabrication of these designer disordered structures is very demanding and required several months of optimization in order to obtain the results reported here. Particular care must be taken how to set up the writing protocol and moreover all writing parameters have to be optimized in order to create mechanically stable structures. We note that precision direct laser writing is almost always operated by sequential writing of lines in 3D (with $840 \times 280$ nm$^2$ cross section in our case) and not by a layer-by-layer process, commonly employed in lower-resolution commercial 3D printing techniques used for rapid prototyping or manufacturing.  Therefore, in our fabrication process particular care must be taken how to set up the writing protocol since there exist no obvious rules how to write a random free-standing network structure at optimal resolution. Moreover it must be ensured that the structure remains mechanically stable in a soft gel photoresist throughout the writing process of about 1 h. This task is further complicated by intrinsic mechanical stresses created upon exposure of the photoresist. The latter leads to substantial deformations if the written lines are not attached within seconds to a mechanically rigid superstructure.  To overcome these challenging problems, encountered in our initial fabrication attempts, we have developed a optimized writing protocol. We first divide the entire volume in cubic sub-volumes of side length roughly $1.5 D$ (where $D$ is the average distance between nearest points in the underlying point pattern). We sort the rods in such a way that i) once all rods belonging to a certain cube are written, we proceed to a neighboring cube, ii) first all cubes closest to the substrate are filled with rods, then the ones lying higher above and so on until the whole network has been written.  Fig. \ref{fig2} displays a representative set of electron micrographs of fabricated structures.  We succeeded to write structures with either a square ($65 \times 65 \mu$m$^2$) or circular (diameter $65 \mu$m) footprint with heights varying between $h=4-12 \mu$m. Structures higher than $4\mu$m were surrounded by a massive wall (Fig. \ref{fig2}) for enhanced mechanical stability  \cite{ref12}.

\subsection{Structural analysis by light scattering} 
\begin{figure}[htbp]
\centering\includegraphics[width=10cm]{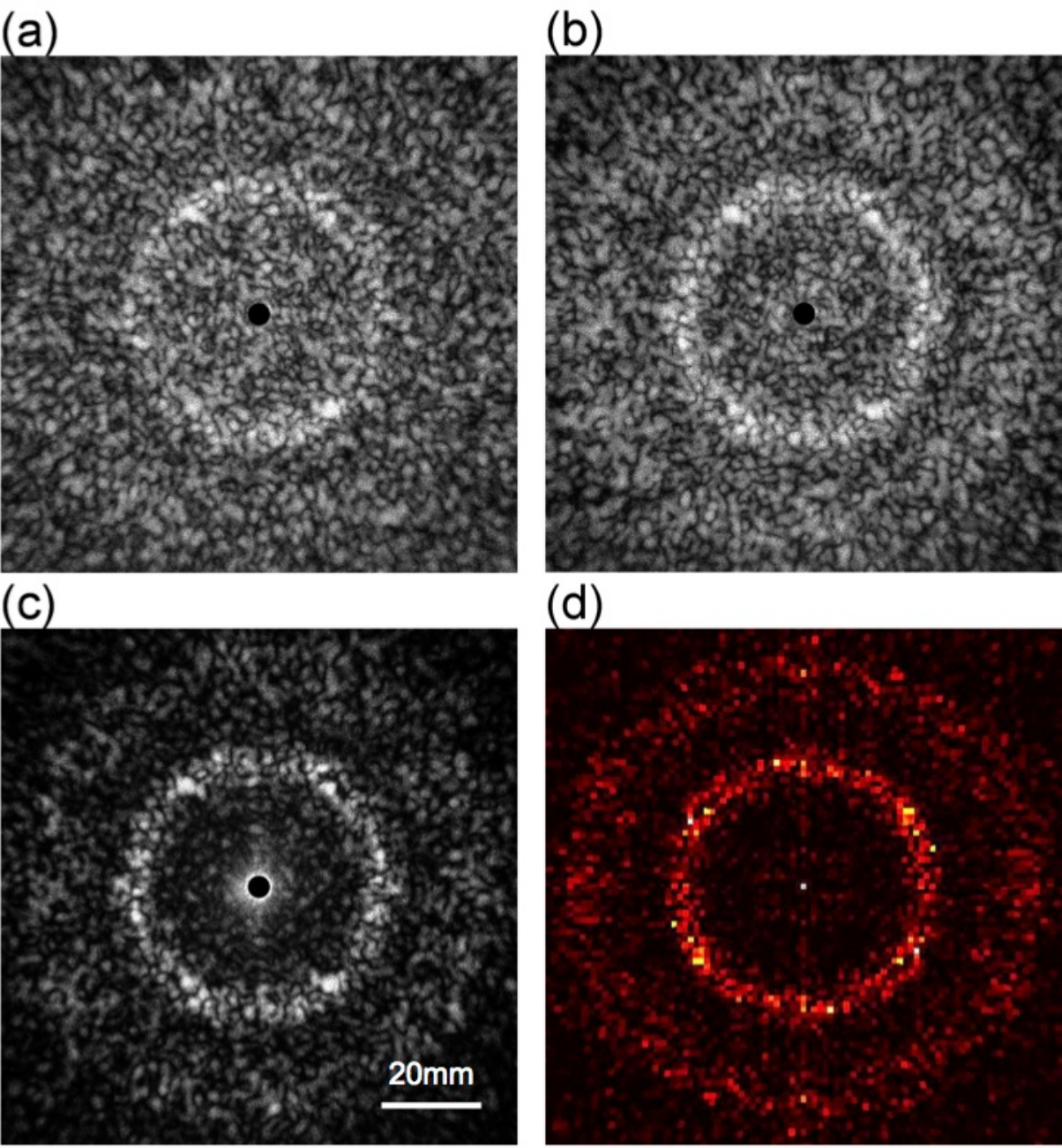}
\caption{Small angle light scattering pattern using a $\lambda=632.8$nm laser. (a),(b),(c), scattering pattern for a structure of thickness $h = 4\mu$m in air, infiltrated with isopropanol ($n=1.377$) and with toluene ($n=1.496$). (d) calculated scattering pattern in the single scattering limit for the same structure based on a discrete dipole approximation (DDA).}
\label{fig3}
\end{figure}

We characterize the properties of our samples by measurements of the scattering patterns using visible light. For the optical characterization of our sub-mm sized samples we have built a small angle scattering instrument consisting of a helium neon laser ($\lambda = 632.8$ nm), a focusing lens (focal length $f=50$ mm), two diaphragms to suppress stray light and a white screen, positioned at a distance of $z= 125$mm from the sample, with a central absorbing beam block. The scattered light pattern is photographed off the white screen using a digital camera. The setup has been calibrated using a small pinhole. Histogram normalization has been applied to all images displayed in Fig. \ref{fig3}.  This procedure represents a linear transformation of an image where the value $P_{in}$ of each pixel is scaled according to \cite{ref16}: $P_{out}=255 (P_{in}-c)/(d-c)$. Here $c$ and $d$ are the $x$-th and $(100-x)$th percentile in the histogram of pixel intensity values. All values $P_{out}$ smaller than zero are set to $0$, while those larger than $255$ are set to $255$. In our case $x=0.1$. Before calculating radial averages of experimental diffraction patterns, several data processing steps were performed. Despite the use of diaphragms we could still observe some stray light contributing to the image. This contribution has to be subtracted from the raw data. To this end we have acquired an image from a bare glass substrate (empty cell) inserted into the laser beam. This image is subsequently subtracted from the raw data. The modulus of the scattering vector $\bf{q}$ is calculated from the radial distance $u$ from the center via the relation $\tan(\theta)=u/z$ and $q=(4¹/\lambda)\sin(\theta/2)$.  These relations apply also for the (toluene or toluene/chlorobenzene) infiltrated structures since for small angles the reduction of the wavelength $\lambda/n$ within the sample is to a good approximation offset by refraction at the flat sample-air interface. Another small correction results from the actual detector acceptance angle being different for each scattering angle $\theta$. This gives rise to a correction \cite{ref19} of the measured data by a factor $\cos(\theta)^{-3}$.
\newline Examples for light scattering data recorded for $h = 4\mu$m structures are shown in Fig. \ref{fig3}. The data clearly reveals a concentric ring profile without Bragg peaks. The pronounced maximum indicates short-range order while the ring-shape is a signature of structural isotropy without any long-range order. Another feature related to disorder is the speckled appearance of the scattering pattern, reminiscent of the laser speckle observed for an arbitrary random structure. Although our first observations confirm the overall picture, the ring pattern obtained in air, Fig. \ref{fig3}(a), appears blurred with a weaker than expected maximum and substantial low angle scattering. Interestingly similar observations have been made for quasicristalline structures \cite{ref17}. In our case we can clearly attribute the blurring to multiple scattering. From the attenuation of the direct laser beam intensity we can estimate a scattering length of $l_s \simeq 3\mu$m, short even compared to the lowest sample studied with $h = 4\mu$m. In order to reduce scattering we infiltrate the sample first with isopropanol ($n=1.377$) and then with toluol ($n=1.496$), which leads to a gradual reduction of multiple scattering and a sharpening of the diffraction ring, Fig. \ref{fig3}(b),(c), while at the same time the ring position remains unchanged. For the latter case the refractive index of the polymeric structure is almost matched and the direct beam is attenuated only by a few percent signaling the absence of multiple scattering. Similar results are obtained for the $h = 8\mu$m structures when using a 1:2 mixture of toluene/chlorobenzene ($n=1.517$) as an index matching fluid (Fig. \ref{fig4} (b).  In both cases a clear and pronounced ring appears in the scattering pattern at $q\sim 2 \pi/d$ and for smaller $q$-values scattering is strongly suppressed. Spatial speckle fluctuations can be largely (although not entirely) suppressed by taking radial averages as shown in Fig. \ref{fig4} .

\subsection{Comparison with numerical calculations} 
\begin{figure}[htbp]
\centering\includegraphics[width=9cm]{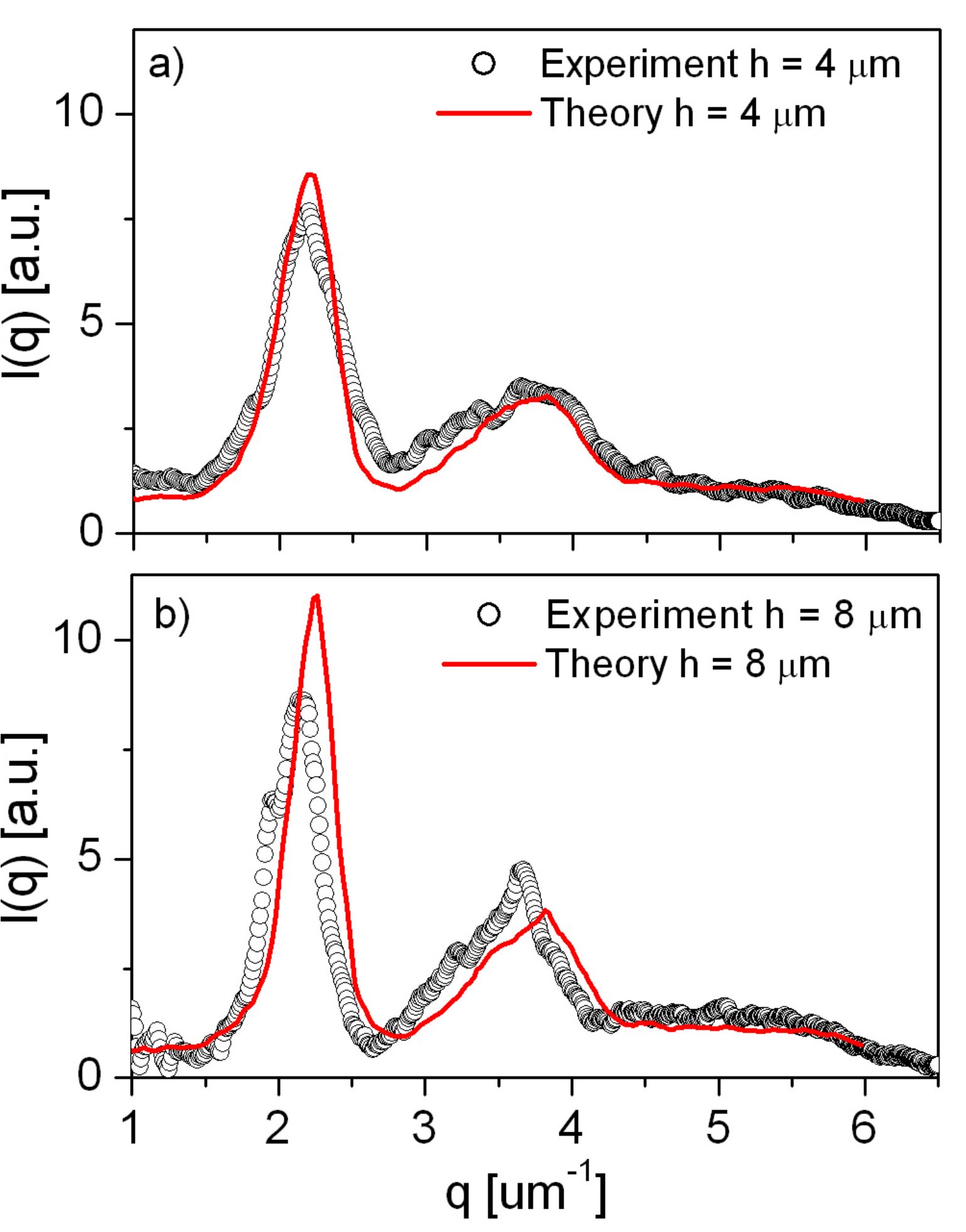}
\caption{Radially averaged scattering intensity. Symbols: experimental data $I(q)$ for $h = 4\mu$m, immersed in toluene ($n=1.496$), and $h = 8\mu$m, immersed in toluene/chlorobenzene 1:2 ($n=1.517$). Solid lines: theoretical calculations for the same heights.}
\label{fig4}
\end{figure}

For a quantitative analysis of the experimental results we numerically solve the scattering problem using a discrete dipole approximation (DDA) in the single scattering limit \cite{ref18}. Given the weak scattering contrast in the toluene- (or toluene/chlorobenzene) infiltrated structures we expect this approximation to hold very well. First, rods coordinates generated in the manner described above are used to create a 3D binary representation of the network (Fig. \ref{fig1} (a)). Namely a sufficiently dense grid is defined and then each grid point is set to 1 if the point belongs to a rod and 0 otherwise (Fig. \ref{fig1} (b)). Using this procedure the lithographic voxel size and ellipsoidal shape (as obtained in our direct laser writing system) can be fully taken into account. Consequently, the binary network represents a good approximation of the true structure as manufactured by laser nanolithography. Following that, a 3D Fast Fourier Transform of this data is calculated. The squared modulus of FFT is proportional to the intensity $I(\bf{q})$ scattered by the structure for a scattering wave vector $\bf{q}=\bf{k}-\bf{k_0}$, where $\bf{k}$ denotes the scattered wave vector and $\bf{k_0}$ the incident wave vector. Lines plotted in Fig. \ref{fig4} have been obtained by averaging over several realizations of a structure of a given height $h$. As shown in Fig. \ref{fig3} (d) the numerical results reproduce both the ring diffraction as well as the superimposed random specular structure. A comparison of the radially averaged numerical data with experiment, Fig.  \ref{fig4} , reveals a very good match with no adjustable parameters except for the absolute scale of intensities.

\section{Discussion} Our optical characterization confirms the high quality of our fabricated samples.  Moreover the results readily show that due to the finite size the optical properties of the material are not yet fully developed. As can be seen in Fig. \ref{fig4} the peak height increases from $h = 4\mu$m to $h = 8\mu$m.  Preliminary numerical results (data not shown) indicate that the peak height saturates for heights larger than 20$\mu m$m.  Moreover both the experimental and numerical results show that the extrapolated values for $S(q\to0)$ are finite.  Due to experimental difficulties accessing wavenumbers close to the primary beam we are unable to clearly distinguish finite-size effects from residual contaminations and experimental artifacts.  Similarly the accuracy of our numerical calculations for small wavenumbers is limited due to finite size effects. Nevertheless, since the seed structure is hyperuniform and since the quality of our polymeric templates is very high, we do believe that the polymeric template possesses the contemplated structural properties and thus should likely give rise to a full PBG when transferred into high-index dielectric. \newline We now turn our attention to the spectral properties of the interconnected network structures. Numerical results predict a broad isotropic bandgap in three dimensions for a refractive index of $n=3.6$ and a volume filling fraction of about $20\%$ \cite{ref8}. In the following we discuss the additional processing steps required in order to obtain such strong photonic properties. Building upon the successful fabrication reported here the following parameters have to be optimized: a) the material refractive index needs to be increased b) the volume filling fraction must be increased to about 20$\%$ and finally the structural features ideally should be further scaled down.  To realize the first point one can rely on well-established procedures reported in the literature. It has been shown that polymeric templates very similar to ours can be transferred into materials such as silicon ($n = 3.6$ for infrared wavelengths) by double inversion retaining the original topology \cite{ref17,Stau10}. Optimal filling fractions have been predicted to be around 20$\%$ while currently, in our case, the polymer content is only around $8\%$. This means that for best results one either has to increase the polymer volume fraction and perform a double inversion or alternatively it should be possible to coat directly the polymeric template with a high index material.  Finally, reducing the structural length scales will require further incremental optimisation of the delicate fabrication process. Such experiments are currently underway in our laboratory.  For the polymeric structures reported here one would expect a mid gap wavelength in air at around $\lambda \sim 4\pi/q_{max} \sim 6\mu$m, about four times larger than typical telecommunication wavelengths of $\lambda \sim 1.5\mu$m \cite{ref8}. Based on preliminary results we expect to be able to reduce the typical length scales by at least a factor of two within the next months. We note that if the feature sizes are scaled down this also favourably affects the filling fraction as long as the size of the laser-writing pen is kept constant.

\section{Summary and conclusion}
We have successfully demonstrated the fabrication of high quality three-dimensional polymeric templates for disordered photonic materials.  Although a number of further processing steps are still required to obtain strong photonic properties, our results already demonstrate the feasibility of creating such complex materials on length scales comparable to optical wavelengths.  As shown previously such polymeric templates can be reliably replicated into materials such as silicon. We thus envision a successful transfer of our polymer templates into a high index material within the near future \cite{Stau10}.   Moreover, in the present study we have shown that the design parameters of polymeric templates can be set precisely using direct laser writing lithography. This in turn will allow rigorous experimental testing of the existing theoretical concepts when applied to the high index replica. 

\section*{Acknowledgments}
We thank Georg Maret and Hui Cao for illuminating discussions and Matteo Molteni and Fabio Ferri for verifying some of our numerical calculations using an independent algorithm. We are grateful to the MPI for polymer research (Mainz) for giving us access to their focused-ion-beam (FIB) instrument and we thank Michael Kappl for help with the experiments. JH acknowledges funding from a Sciex Swiss Research Fellowship No. 10.030. We thank the Swiss National Science Foundation (projects 132736 and 128729) and the Adolphe Merkle Foundation for financial support.

\end{document}